\documentclass[letterpaper,12pt]{article}
\usepackage{lscape}
\usepackage{graphicx}
\usepackage{amsmath}
\usepackage{amssymb}
\usepackage{hyperref}
\usepackage{graphicx,epic}

\usepackage{multicol,color}

\definecolor{darkred}{rgb}{0.5,0,0.5}
\hypersetup{pdfborder={0 0 0},colorlinks=true,urlcolor=darkred,citecolor=blue,linkcolor=darkred}

\makeatletter

\@addtoreset{equation}{section}
\makeatother

\newcommand{\nn}{\nonumber}
\newcommand{\lp}{\left(}
\newcommand{\rp}{\right)}

\newcommand{\ints}{\mathbb{Z}}
\newcommand{\reals}{\mathbb{R}}

\newcommand{\gs}{g_{\rm s}}
\newcommand{\cE}{\mathcal{E}}
\newcommand{\cA}{\mathcal{A}}

\newcommand{\scp}{\sigma}
\newcommand{\newphi}{\varphi}
\newcommand{\cF}{\mathcal{F}}

\begin{document}
\thispagestyle{empty}

{\hfill{\tt AEI-2011-063}}
\vspace{20mm}

\begin{center} {\bf \LARGE Counting supersymmetric branes}

\vspace{15mm}

Axel Kleinschmidt

\footnotesize
\vspace{9mm}

{\em Max Planck Institute for Gravitational Physics, Albert Einstein Institute\\
Am M\"uhlenberg 1, 14476 Potsdam, Germany\\
 \& \\
 International Solvay Institutes\\ Campus Plaine C.P. 231, Boulevard du
Triomphe, 1050 Bruxelles, 
Belgium}

\vspace{4mm}

\vspace {7mm}

\parbox{130mm}{

\noindent \footnotesize Maximal supergravity solutions are revisited and classified, with particular emphasis on objects of co-dimension at most two. This class of solutions includes branes whose tension scales with $\gs^{-\scp}$ for $\scp>2$. We present a group theory derivation of the counting of these objects based on the corresponding tensor hierarchies derived from $E_{11}$ and discrete T- and U-duality transformations. This provides a rationale for the wrapping rules that were recently discussed for $\scp\leq 3$ in the literature and extends them. Explicit supergravity solutions that give rise to co-dimension two branes are constructed and analysed.}

\vspace{15mm}

{\em Dedicated to the memory of Laurent Houart}

\end{center}

\setcounter{equation}{0}
\setcounter{page}{0}

\newpage

\section{Introduction}
\label{sec:intro}

It is by now well-established that maximal supergravity in $D\geq 3$ dimensions admits a systematic hierarchy of $p$-form potentials with $0\leq p \leq D$. This structure has been termed tensor hierarchy~\cite{deWit:2005hv,deWit:2008ta,deWit:2008gc} and has been derived from the properties of gauged maximal supergravity theories. Tantalizingly, the tensor hierarchy derived in this way agrees very well with a `spectral analysis' of the Kac--Moody Lie algebra $E_{11}$~\cite{Riccioni:2007au,Bergshoeff:2007qi}.\footnote{There are small (potential) differences in $D=3$ that we will argue in footnote~\ref{fn:D3} are not important for the purposes of this paper.} The compatibility of the tensor hierarchy with supersymmetry was also checked in many cases, most notably the occurence of ten-forms in type IIB supergravity in $D=10$~\cite{Kleinschmidt:2003mf,Bergshoeff:2010mv}. The various $p$-forms that appear in the tensor hierarchy transform under the continuous hidden symmetry groups $E_{11-D}(\reals)$. For this paper we will use the tensor hierarchy as predicted by $E_{11}$ and summarized in table~\ref{tab:TH}.

In supergravity, there typically exist relatively simple solutions that couple to a $p$-form potential. These are called $(p-1)$ branes and have a $p$-dimensional world-volume and a string frame tension with fixed scaling with respect to the string coupling constant $\gs$. For example, in the line for IIA in table~\ref{tab:TH} one can see the standard supersymmetric D$0$, F$1$, D$2$, D$4$, NS$5$, D$6$ and D$8$ branes of the theory. In addition, there are a $p=8$ potential and two $p=10$ potentials. The question which of these couple to {\em supersymmetric} branes has been answered for example in~\cite{Bergshoeff:2006qw,Bergshoeff:2010mv} where it was found that none of them gives rise to a supersymmetric brane in type IIA.

Looking at table~\ref{tab:TH}, the question arises more generally: {\em Out of the $p$-form potentials in a given dimension $D$, which ones are sources for supersymmetric branes of maximal supergravity and what is the scaling of their tension with $\gs$?} This is the main question we will answer fully and completely using $E_{11}$ in this paper. We note that quite a number of results in this direction have already been obtained in the recent past~\cite{Bergshoeff:2010xc,Bergshoeff:2011zk,Bergshoeff:2011mh,Bergshoeff:2011ee}.  The analysis rested on analysing the structure of the Wess--Zumino terms as a necessary criterion for supersymmetry and resulted in a number of wrapping rules that permit the derivation of the number of supersymmetric solitons of a given scaling $\gs^{-\scp}$ for $\scp=0,1,2,3$ from the known branes in IIA and IIB. An additional outcome of this analysis was that, in order to get the right number of solutions, it is not sufficient to just study the dimensional reduction of the IIA/IIB branes but a supplemental `inflow' of objects was needed for $\scp\geq 2$ from so-called 

\begin{landscape}
\begin{table}
\centering
\scalebox{.9}{{\small\begin{tabular}{c||cccccccccc}
\hline\hline&&&&&&&&&\\[-2mm]
$D \backslash p$ & $1$ & $2$ & $3$ & $4$ & $5$ & $6$ & $7$ & $8$ & $9$ & $10$\\[2mm]
\hline\hline&&&&&&&&&\\[-2mm]
IIA & ${\bf 1}$ & ${\bf 1}$ & ${\bf 1}$ & & ${\bf 1}$ & ${\bf 1}$ & ${\bf 1}$ & ${\bf 1}$ & ${\bf 1}$ & $2\times{\bf 1}$\\[3mm]
IIB &  & ${\bf 2}$ & & ${\bf 1}$ & & ${\bf 2}$ && ${\bf 3}$ && ${\bf 4} \oplus {\bf 2}$\\[3mm]
\hline&&&&&&&&&\\[-2mm]
$9$ & ${\bf 2}\oplus {\bf 1}$ & ${\bf 2}$ & ${\bf 1}$ & ${\bf 1}$ & ${\bf 2}$ & ${\bf 2}\oplus{\bf 1}$ & ${\bf 3}\oplus {\bf 1}$ & ${\bf 3}\oplus {\bf 2}$ & ${\bf 4}\oplus2\times{\bf 2}$\\[3mm]
\hline&&&&&&&&&\\[-2mm]
$8$ & $({\bf 3},{\bf 2})$ & $({\bf \bar{3}},{\bf 1})$ & $({\bf 1},{\bf 2})$ &  $({\bf 3},{\bf 1})$ &  $({\bf \bar{3}},{\bf 2})$ &  $({\bf 8},{\bf 1})\oplus ({\bf 1},{\bf 3})$ &  $ ({\bf \bar{6}},{\bf 2})\oplus({\bf 3},{\bf 2})$ &  $\begin{array}{c}({\bf 15},{\bf 1})\oplus ({\bf \bar{3}},{\bf 3})\\\oplus 2\times ({\bf \bar{3}},{\bf 1})\end{array}$&\\[3mm]
\hline&&&&&&&&&\\[-2mm]
$7$ & ${\bf \overline{10}}$ & ${\bf 5}$ & ${\bf \bar{5}}$ & ${\bf 10}$ & ${\bf 24}$ & ${\bf 40}\oplus{\bf \overline{15}}$  & ${\bf \overline{70}}\oplus{\bf \overline{45}}\oplus{\bf 5}$&& \\[3mm]
\hline&&&&&&&&&\\[-2mm]
$6$ &  ${\bf 16_c}$ & ${\bf 10}$ & ${\bf 16_s}$ & ${\bf 45}$ & ${\bf 144_s}$ & $\begin{array}{c}{\bf 320}\oplus{\bf 126_+}\\\oplus{\bf 10}\end{array}$&&&\\[3mm]
\hline&&&&&&&&&\\[-2mm]
$5$ & ${\bf \overline{27}}$ & ${\bf 27}$ & ${\bf 78}$ & ${\bf 351}$ & ${\bf 1728}\oplus {\bf 27}$ &&&&\\[3mm]
\hline&&&&&&&&&\\[-2mm]
$4$ & ${\bf 56}$ & ${\bf 133}$ & ${\bf 912}$ & ${\bf 8645}\oplus{\bf 133}$ &&&&&\\[3mm]
\hline&&&&&&&&&\\[-2mm]
$3$ & ${\bf 248}$ & ${\bf 3875}\oplus {\bf 1}$ & $\begin{array}{c}{\bf 147250}\\\oplus{\bf 3875}\oplus{\bf 248}\end{array}$&&&&&&\\[3mm]
\hline\hline
\end{tabular}}}
{\caption{\label{tab:TH}\sl The tensor hierarchy of $p$-forms for $3\leq D \leq 10$ as predicted by $E_{11}$. For type IIA, there is no non-abelian U-duality, for type IIB it is $SL(2,\reals)$. We omit the scalars that form the (non-linear) coset $E_{11-D}/K(E_{11-D})$ with $K(E_{11-D})$ the maximal compact subgroup. In $D=8$, we take $E_3=SL(3)\times SL(2)$. See also~\cite{Riccioni:2007au,Bergshoeff:2007qi}.}}
\end{table}
\end{landscape}

\noindent generalised Kaluza--Klein monopoles. This only happens for branes of co-dimension at most two. Branes of co-dimension less than two couple to $p$-forms with $p=D-1$ or $p=D$ and hence not to fields with local degrees of freedom.

That not all $p$-forms give rise to supersymmetric solutions can be seen for example in the type IIB case. There one obtains an $SL(2,\reals)$ triplet of $8$-forms~\cite{Kleinschmidt:2003mf,Bergshoeff:2005ac} that would na\"ively suggest a triplet of supersymmetric $7$-branes. One supersymmetric $7$-brane is certainly the D$7$ (with $\scp=1$), another one is its S-dual partner (with $\scp=3$). However, it turns out that one cannot associate a supersymmetric $7$-brane with the third $8$-form potential~\cite{Bergshoeff:2006gs,Bergshoeff:2006jj,Bergshoeff:2011zk}. The reason is basically that these are dual to only the two scalars $(\phi,\chi)$ parametrising the $SL(2,\reals)/SO(2)$ scalar manifold. This kind of reasoning will not apply to $p$-forms with $p=D-1$ or $p=D$ in $D$ space-time dimensions since the corresponding forms are not dual to any of the physical fields.\footnote{The $(D-1)$-forms correspond to possible (gauge) deformations of maximal supergravity, the $D$-forms are related to constraints on these deformations; this will not be of importance in this paper.} U-duality is compatible with supersymmetry, so all the members of the U-duality orbit of a given supersymmetric brane will be supersymmetric. This is what happens in the example above: the third potential is not in the orbit of the D$7$-brane.

With this reasoning we could already determine a lower bound on the number of supersymmetric $p$-branes in $D$ space-time dimensions by looking at the size of the U-duality orbits of the standard branes.\footnote{The counting of supersymmetric branes presented here only deals with `pure' or `elementary' branes and does not consider intersecting branes nor dyonic~\cite{Izquierdo:1995ms} or more complicated bound states. These can be also analysed by algebraic methods~\cite{Englert:2003py,Englert:2004it,Cook:2009ri,Houart:2009ya}.}  
We claim that this already gives the full correct answer rather than just a lower bound. As will be explained in more detail below, the reason is that from an $E_{11}$ point of view, this orbit corresponds entirely to {\em real} roots of the algebra, whereas all other orbits correspond to {\em imaginary} roots. That the solutions associated to imaginary roots are not supersymmetric was verified in detail in the null case in~\cite{Houart:2011sk}.\footnote{The difference between real and imaginary also played an important role in~\cite{Brown:2004jb}.} What has to be done therefore from an $E_{11}$ logic is to select among the potentials of table~\ref{tab:TH} those that correspond to real roots. This can be done since one knows for all the potentials the corresponding roots. What one arrives at is table~\ref{tab:SB}. For consistency we note that U-duality preserves the norm.\footnote{\label{fn:D3}The difference in the $3$-form spectrum in $D=3$ in the tensor hierarchy analysis of~\cite{deWit:2008ta} and the $E_{11}$ analysis of~\cite{Riccioni:2007au,Bergshoeff:2007qi} consists of the ${\bf 248}$ of $E_8$ in table~\ref{tab:TH}. It lies entirely in the imaginary root sector and so does not source any supersymmetric branes. Its presence or not does not alter table~\ref{tab:SB}.}

\begin{table}[t!]
\centering
\begin{tabular}{c||cccccccccc}
\hline\hline&&&&&&&&&&\\[-2mm]
$D\backslash p$ & $0$ & $1$ & $2$ & $3$ & $4$ & $5$ & $6$ & $7$ & $8$ & $9$ \\[2mm]
\hline\hline\\[-3mm]
IIA & $1$ & $1$ & $1$ & & $1$ & $1$ & $1$ && $1$ & \\[2mm]
IIB & & $2$ && $1$ && $2$ && $2$ && $2$\\[2mm]
\hline&&&&&&&&&&\\[-2mm]
$9$ & 3 & 2 & 1 & 1 & 2 & 3 & 2 & 2 & 2&\\[2mm]
\hline&&&&&&&&&&\\[-2mm]
$8$ & 6 &3 & 2 & 3 & 6 & 8 & 6 & 6&&\\[2mm]
\hline&&&&&&&&&&\\[-2mm]
$7$ & 10 & 5 & 5 & 10 & 20 & 25 & 20 &&&\\[2mm]
\hline&&&&&&&&&&\\[-2mm]
$6$ & 16 & 10 & 16 & 40 & 80 & 96 &&&&\\[2mm]
\hline&&&&&&&&&&\\[-2mm]
$5$ & 27 & 27 & 72 & 216 & 432&&&&&\\[2mm]
\hline&&&&&&&&&&\\[-2mm]
$4$ & 56 & 126 & 576 & 2016&&&&&\\[2mm]
\hline&&&&&&&&&&\\[-2mm]
$3$ & 240 & 2160 & 17280&&&&&&& \\[2mm]
\hline\hline

\end{tabular}
{\caption{\label{tab:SB}\sl Total number of supersymmetric $p$-branes of maximal supergravity in $D$ space-time dimensions as predicted by $E_{11}$. Note that $p$ is shifted by one compared to table~\ref{tab:TH} as it now labels the $p$-branes (coupling to a $(p+1)$-form potential). The table contains all $\gs$ scalings together.}}
\end{table}

In the table, all possible $\gs$ scalings are still grouped together but it is desirable to differentiate the different types of non-perturbative behaviour in terms of string frame tensions. This can be done by recalling that U-duality is composed out of T- and S-duality, see for instance.~\cite{Hull:1994ys,Obers:1998fb}. T-duality does not change the power of $\gs$~\cite{Obers:1998fb} and therefore one needs to consider the breaking 
\begin{align}
\label{utalpha}
E_{11-D}(\reals) \supset SO(10-D,10-D;\reals)\times \reals_+.
\end{align}
The weight under $\reals_+$ corresponds to $\scp$ when normalized correctly. We will discuss this breaking in more detail below and there will give tables counting the supersymmetric objects for all values of $p$, $D$ and $\scp$. Our results represent the first complete count of supersymmetric branes in all dimensions $D\geq 3$ and agree with completely independent derivations where available.

Viewing the supersymmetric branes as associated to real roots of $E_{11}$ makes it very simple to follow them through the various dimensions and obtain higher-dimensional origins of the generalised KK-monopoles and similar `obscure' non-standard objects necessary for the correct counting. 

Besides counting the solutions, it is of interest to construct them. This can be done by performing U-duality transformations on known solutions and in this way generalised KK monopoles of co-dimension two have already been constructed in the past~\cite{LozanoTellechea:2000mc,Englert:2007qb}. These solutions require a fair number of isometry directions when viewed as ten-dimensional solutions because one needs to perform many T-duality transformations on the known solutions. 

We will not address here the construction of supersymmetric generalised Kaluza--Klein monopole solutions with co-dimension less than two but leave this to future work. The reason is that they require the extension of the U-duality transformation rules to deformed supergravity, most notably to massive type IIA supergravity where the D$8$-brane is already a solution only of the mass deformed theory~\cite{Polchinski:1995mt}. Some solutions can be constructed without such generalizations of e.g.~\cite{Bergshoeff:1996ui} and which ones can easily be decided by studying the corresponding $E_{11}$ root vectors.

The rest of this note starts by giving a more detailed description of the $E_{11}$ derivation of the counting of branes and tables for all values of $p$, $D$ and $\scp$ in section~\ref{sec:e11sec}. There we will also discuss the higher-dimensional ($D=11$) origin of the solutions. In section~\ref{sec:sols}, we explicitly construct local solutions using T-duality transformations. Appendix~\ref{app:IIAIIBorigin} gives the type IIA and type IIB origin of all the branes and appendix~\ref{app:conv} contains our conventions.

\section{$E_{11}$ derivation}
\label{sec:e11sec}

We start this section by recalling the usefulness of $E_{11}$~\cite{West:2001as} as a bookkeeping device for the field content of maximal supergravity in various space-time dimensions~\cite{Kleinschmidt:2003mf,Riccioni:2007au,Bergshoeff:2007qi}.

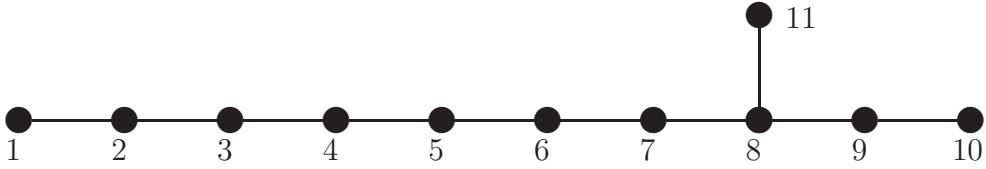
\begin{figure}[t!]
\centering
\begin{picture}(380,60)
\put(5,-5){$1$}
\put(45,-5){$2$}
\put(85,-5){$3$}
\put(125,-5){$4$}
\put(165,-5){$5$}
\put(205,-5){$6$}
\put(245,-5){$7$}
\put(285,-5){$8$}
\put(325,-5){$9$}
\put(363,-5){$10$}
\put(300,45){$11$}
\thicklines
\multiput(10,10)(40,0){10}{\circle*{10}}
\multiput(15,10)(40,0){9}{\line(1,0){30}}
\put(290,50){\circle*{10}} \put(290,15){\line(0,1){30}}
\put(290,10){\circle*{10}}
\end{picture}
{\caption{\label{fig:e11dynk}\sl Dynkin diagram of $E_{11}$ with labelling of nodes.}}
\end{figure}

The Dynkin diagram of $E_{11}$ is given in figure~\ref{fig:e11dynk}. If one is interested in maximal supergravity in $D$ dimensions one has to `cut' the diagram at node $D$.  To the left one will have a diagram of type $A_{D-1}$, corresponding to the so-called `gravity line'. To the right of the cut one obtains the diagram of the hidden symmetry $E_{11-D}$.\footnote{For $D=9$, one has to also remove node $11$, for $D=10$, there are two choices, corresponding to type IIA and type IIB~\cite{Kleinschmidt:2003mf}.} In other words one has a decomposition\footnote{The enhancement from $SL(D,\reals)\cong A_{D-1}$ to $GL(D,\reals)$ is due to the additional Cartan generator of the node where one cut the diagram.}
\begin{align}
\label{maxD}
E_{11} \supset GL(D,\reals) \times E_{11-D}.
\end{align}

One obtains the spectrum of fields in $D$ space-time dimensions by decomposing the adjoint representation of $E_{11}$ under this subgroup. As one deals with the adjoint representation one can label the elements by the roots $\alpha$ of the algebra. Any such root can be expanded in a basis of simple roots $\alpha_i$ as
\begin{align}
\alpha = \sum_{i=1}^{11} m^i \alpha_i.
\end{align}
As $E_{11}$ is an infinite-dimensional Kac--Moody algebra, $\alpha$ can be either real or imaginary, depending on whether its norm $|\alpha|^2$, computed in the standard Killing metric, is greater than zero or not. The simplest example of an imaginary root is the affine null root
\begin{align}
\label{eq:nullrt}
\delta = \alpha_3 + 2 \alpha_4 + 3\alpha_5 + 4\alpha_6+ 5\alpha_7 + 6\alpha_8 + 4\alpha_9 +2\alpha_{10} + 3 \alpha_{11}.
\end{align}
Imaginary roots can have multiplicities; the multiplicity of $\delta$ is equal to 8. Real roots always have multiplicity equal to $1$.

The result of the decomposition of the adjoint of $E_{11}$ under (\ref{maxD})  can be ordered by `level', that is by the number of times the deleted simple node $\alpha_D$ appears in the expansion. The results of this decomposition at low levels for all $D$ can be found in~\cite{Kleinschmidt:2003mf,Riccioni:2007au,Bergshoeff:2007qi}. In this way one can associate to any given field in maximal supergravity a root $\alpha$ {\em et vice versa}.\footnote{Since only Borel elements are associated with fields in a non-linear realization, we always have $\alpha>0$. Due to our restriction to real roots in the following we are not displaying multiplicity indices.}

\begin{table}
\centering
\begin{tabular}{c||cccccccccc}
\hline\hline&&&&&&&&&&\\[-4mm]
$D\backslash p$ & $0$ & $1$ & $2$ & $3$ & $4$ & $5$ & $6$ & $7$ & $8$ & $9$ \\
\hline\hline\\[-4mm]
IIA & & 1 &  & & & & &&  & \\[1mm]
IIB & & 1 &&  &&  &&  && \\
\hline&&&&&&&&&&\\[-4mm]
$9$ & 2 & 1 &  & &  &  &  & & &\\
\hline&&&&&&&&&&\\[-4mm]
$8$ & 4 &1 &  &  &  &  &  & &&\\
\hline&&&&&&&&&&\\[-4mm]
$7$ & 6 & 1 &  &  & & &  &&&\\
\hline&&&&&&&&&&\\[-4mm]
$6$ & 8 & 1 &  &  &  &  &&&&\\
\hline&&&&&&&&&&\\[-4mm]
$5$ & 10 & 1 & &  & &&&&&\\
\hline&&&&&&&&&&\\[-4mm]
$4$ & 12 & 1 &  & &&&&&\\
\hline&&&&&&&&&&\\[-4mm]
$3$ & 14 & 1 & &&&&&&& \\
\hline\hline

\end{tabular}
{\caption{\label{tab:scp0}\sl Supersymmetric $p$-branes in $D$ dimensions that scale like $\gs^{-0}$, i.e., $\scp=0$. The table agrees with that of~\cite{Bergshoeff:2011mh} and is in accord with the `fundamental wrapping rule'. The table contains the fundamental string and its reductions as well reductions of KK momentum states.}}
\end{table}

As explained in the introduction, we are interested in the breaking of U-duality to T-duality.
The U-duality group $E_{11-D}$ is constructed from the nodes from $D+1$ up to $11$; the $SO(10-D,10-D)$ T-duality subgroup is obtained by ignoring the node $10$ of the $E_{11-D}$ diagram. The $\reals_+$ factor in (\ref{utalpha}) is therefore associated with the scaling under node $10$. More precisely, we have~\cite{Kleinschmidt:2003mf}\footnote{See also~\cite{Cook:2008bi}.}
\begin{align}
\scp = m^{10},
\end{align}
i.e., the string frame tension dependence $\scp$ is the eigenvalue under the fundamental weight generator $\lambda_{10}$ acting on a given generator of $E_{11}$.

\begin{table}
\centering
\begin{tabular}{c||cccccccccc}
\hline\hline&&&&&&&&&&\\[-4mm]
$D\backslash p$ & $0$ & $1$ & $2$ & $3$ & $4$ & $5$ & $6$ & $7$ & $8$ & $9$ \\
\hline\hline\\[-4mm]
IIA &1 &  & 1 & & 1& &1 &&1  & \\[1mm]
IIB & & 1 && 1 && 1 && 1 && 1\\
\hline&&&&&&&&&&\\[-4mm]
$9$ & 1 & 1 & 1 & 1 &  1 &  1 & 1 & 1 & 1&\\
\hline&&&&&&&&&&\\[-4mm]
$8$ & 2 & 2 & 2 & 2 &  2 & 2 & 2 & 2 &&\\
\hline&&&&&&&&&&\\[-4mm]
$7$ & 4 & 4 &  4 & 4  & 4 & 4 & 4 & &&\\
\hline&&&&&&&&&&\\[-4mm]
$6$ & 8 & 8 &  8 & 8  &  8 &  8 &  &&&\\
\hline&&&&&&&&&&\\[-4mm]
$5$ & 16 & 16 & 16 & 16  & 16 &  &&&&\\
\hline&&&&&&&&&&\\[-4mm]
$4$ & 32 & 32 & 32  & 32 &  &&&&\\
\hline&&&&&&&&&&\\[-4mm]
$3$ & 64 & 64 & 64 &&&&&&& \\
\hline\hline

\end{tabular}
{\caption{\label{tab:scp1}\sl Supersymmetric $p$-branes in $D$ dimensions that scale like $\gs^{-1}$, i.e., $\scp=1$.  The table agrees with that of~\cite{Bergshoeff:2011mh} and is in accord with the `D-brane wrapping rule'. The table contains the usual D-brane spectrum.}}
\end{table}

Let us view this in an example from type IIB. According to~\cite{Kleinschmidt:2003mf}, there are three $8$-form potentials, these correspond to the three roots
\begin{align}
\delta-\alpha_{10},\quad \delta \quad\text{and}\quad \delta+\alpha_{10}.
\end{align}
Referring back to (\ref{eq:nullrt})  the scaling factors of the three potentials are therefore
\begin{align}
\scp=1,\quad \scp=2\quad\text{and}\quad\scp=3.
\end{align}
The first potential is the one that couples to the D$7$ of type IIB, the last one is its S-dual. Both the corresponding roots are real. The middle potential with $\scp=2$ corresponds to the null root $\delta$ and does not give rise to a supersymmetric solution~\cite{Houart:2011sk}. Even though the three potentials together form the ${\bf 3}$ of $SL(2,\reals)$ one cannot map the $\scp=1$ potential fully on $\scp=2$ potential.

Now, we have all the material at hand to find the number of supersymmetric $p$-brane solutions of a given scaling $\scp$ in $D$ space-time dimensions. It is straight-forward to implement this on the computer and we merely give the results in table form in tables~\ref{tab:scp0} to~\ref{tab:scpn}.  One computes all real $E_{11}$ roots (up to height $161$) and selects all those that correspond to antisymmetric tensor, groups them by rank, weight under $\lambda_{10}$ and dimension.\footnote{The identification of the root associated with a given potential can for example be done with the tables of~\cite{Kleinschmidt:2003mf} or with the help of the simpLie software of~\cite{Bergshoeff:2007qi}.} Alternatively, one can take the tensor hierarchy representations of table~\ref{tab:TH}, compute the Weyl orbits of the dominant (real) weights (this gives table~\ref{tab:SB}) and then branch these Weyl orbits under the T-duality Weyl group.\footnote{In some cases there are contributions from more than one irreducible U-duality representation.} Note that we have chosen to display the tables with $\scp$ held fix. In this way, one does not immediately see the feature that for  $D=3,4,6,10$, the numbers of $p$-branes with $p$ fixed and varying $\scp$ are always symmetric, reflecting the underlying $\ints_2$ S-duality. For the remaining values of $D$ there is of course also a $\ints_2$ S-duality that acts on the representations but in these cases the branched orbits arrange themselves asymmetrically. The restriction to real roots we have employed here also follows from the fact that U-duality (the Weyl group) preserves the norm of the roots.

\begin{table}
\centering
\begin{tabular}{c||cccccccccc}
\hline\hline&&&&&&&&&&\\[-4mm]
$D\backslash p$ & $0$ & $1$ & $2$ & $3$ & $4$ & $5$ & $6$ & $7$ & $8$ & $9$ \\
\hline\hline\\[-4mm]
IIA & &  &  & & & 1& &&  & \\[1mm]
IIB & &  &&  && 1 && && \\
\hline&&&&&&&&&&\\[-4mm]
$9$ &  &  &  & &  1 &  2 &  &  & &\\
\hline&&&&&&&&&&\\[-4mm]
$8$ &  &  &  & 1 &  4 & 4 &  &  &&\\
\hline&&&&&&&&&&\\[-4mm]
$7$ &  &  &  1 & 6  &12 & 8 &  & &&\\
\hline&&&&&&&&&&\\[-4mm]
$6$ &  & 1 &  8 & 24  &  32 &  16 &  &&&\\
\hline&&&&&&&&&&\\[-4mm]
$5$ & 1 & 10 & 40 & 80  & 80 &  &&&&\\
\hline&&&&&&&&&&\\[-4mm]
$4$ & 12 & 60 & 160  & 240 &  &&&&\\
\hline&&&&&&&&&&\\[-4mm]
$3$ & 84 & 280 & 560 &&&&&&& \\
\hline\hline

\end{tabular}
{\caption{\label{tab:scp2}\sl Supersymmetric $p$-branes in $D$ dimensions that scale like $\gs^{-2}$, i.e., $\scp=2$.  The table agrees with that of~\cite{Bergshoeff:2011mh} and is in accord with the `dual wrapping rule'. This table contains the NS-branes and reductions of (generalised) Kaluza--Klein monopoles.}}
\end{table}

\begin{table}
\centering
\begin{tabular}{c||cccccccccc}
\hline\hline&&&&&&&&&&\\[-4mm]
$D\backslash p$ & $0$ & $1$ & $2$ & $3$ & $4$ & $5$ & $6$ & $7$ & $8$ & $9$ \\
\hline\hline\\[-4mm]
IIA & &  &  & & & & &&  & \\[1mm]
IIB & &  &&  &&  && 1&& \\
\hline&&&&&&&&&&\\[-4mm]
$9$ &  &  &  & &   &   &  1&1  & &\\
\hline&&&&&&&&&&\\[-4mm]
$8$ &  &  &  &  &   & 2 & 4 & 2 &&\\
\hline&&&&&&&&&&\\[-4mm]
$7$ &  &  &   &   &4 & 12 & 12 & &&\\
\hline&&&&&&&&&&\\[-4mm]
$6$ &  &  &   & 8  &  32 &  48 &  &&&\\
\hline&&&&&&&&&&\\[-4mm]
$5$ &  &  & 16 & 80  & 160 &  &&&&\\
\hline&&&&&&&&&&\\[-4mm]
$4$ &  & 32 & 192  & 480 &  &&&&\\
\hline&&&&&&&&&&\\[-4mm]
$3$ & 64 & 448 & 1344 &&&&&&& \\
\hline\hline

\end{tabular}
{\caption{\label{tab:scp3}\sl Supersymmetric $p$-branes in $D$ dimensions that scale like $\gs^{-3}$, i.e., $\scp=3$.  The table agrees with that of~\cite{Bergshoeff:2011ee} and is in accord with the `exotic wrapping rule'.}}
\end{table}

The higher-dimensional origin of a given object can also be recovered easily using the root $\alpha$ corresponding to the potential and its meaning in the various dimensions. We have performed this for the higher dimensional origin in type IIA, type IIB and M-theory.

It turns out that there is quite a number of mixed symmetry tensors that occur in this analysis. Here, we only give the list of tensors in $D=11$ (M-theory) that contribute to the counting and refer to appendix~\ref{app:IIAIIBorigin} for type IIA and type IIB. We denote by $T_{(M_1,M_2,M_3,\ldots)}$ an irreducible tensor of $GL(11,\reals)$ with Young tableau consisting of columns of lengths $M_1$, $M_2$, etc.. We restrict to $11\geq M_1\geq M_2 \geq \ldots$, and columns of length $11$ are needed to keep track of the overall weight. In this notation, the three-form of $D=11$ supergravity is therefore written as $T_{(3)}$. The graviton is in the adjoint of $GL(11,\reals)$ and is listed as $T_{(10,1)}$ below.

\begin{table}
\centering
\begin{tabular}{c||cccccccccc}
\hline\hline&&&&&&&&&&\\[-4mm]
$D\backslash p$ & $0$ & $1$ & $2$ & $3$ & $4$ & $5$ & $6$ & $7$ & $8$ & $9$\\
\hline\hline\\[-4mm]
IIA & &  &  & & & & &&  & \\[1mm]
IIB & &  &&  &&  && && 1\\
\hline&&&&&&&&&&\\[-4mm]
$9$ &  &  &  & &   &   &  &  &1 &\\
\hline&&&&&&&&&&\\[-4mm]
$8$ &  &  &  &  &   &  &  & 2 &&\\
\hline&&&&&&&&&&\\[-4mm]
$7$ &  &  &   &   & & 1 & 4 & &&\\
\hline&&&&&&&&&&\\[-4mm]
$6$ &  &  &   &  &  8 &  16 &  &&&\\
\hline&&&&&&&&&&\\[-4mm]
$5$ &  &  &  & 40  & 96 &  &&&&\\
\hline&&&&&&&&&&\\[-4mm]
$4$ &  & 1 & 160  & 512 &  &&&&\\
\hline&&&&&&&&&&\\[-4mm]
$3$ & 14 & 574 & 2304 &&&&&&& \\
\hline\hline
\end{tabular}
{\caption{\label{tab:scp4}\sl Supersymmetric $p$-branes in $D$ dimensions that scale like $\gs^{-4}$, i.e., $\scp=4$. There is no obvious wrapping rule associated with this table.}}
\end{table}

\begin{table}
\centering
\begin{tabular}{c||cccccccccc}
\hline\hline&&&&&&&&&&\\[-4mm]
$D\backslash p$ & $0$ & $1$ & $2$ & $3$ & $4$ & $5$ & $6$ & $7$ & $8$ & $9$\\
\hline\hline\\[-4mm]
$6$ &  &  &   &  &   &  8 &  &&&\\
\hline&&&&&&&&&&\\[-4mm]
$5$ &  &  &  &   & 80 &  &&&&\\
\hline&&&&&&&&&&\\[-4mm]
$4$ &  &  & 32  & 480 &  &&&&\\
\hline&&&&&&&&&&\\[-4mm]
$3$ &  & 448 & 2688 &&&&&&& \\
\hline\hline

\end{tabular}
{\caption{\label{tab:scp5}\sl Supersymmetric $p$-branes in $D$ dimensions that scale like $\gs^{-5}$, i.e., $\scp=5$. Dimensions in which there are no branes have been eliminated from the table.}}
\end{table}

\begin{table}
\centering
\begin{tabular}{c||cccccccccc}
\hline\hline&&&&&&&&&&\\[-4mm]
$D\backslash p$ & $0$ & $1$ & $2$ & $3$ & $4$ & $5$ & $6$ & $7$ & $8$ & $9$\\
\hline\hline\\[-4mm]
$4$ &  &  &  & 240 &  &&&&\\
\hline&&&&&&&&&&\\[-4mm]
$3$ &  & 280 & 3360 &&&&&&& \\
\hline\hline

\end{tabular}
{\caption{\label{tab:scp6}\sl Supersymmetric $p$-branes in $D$ dimensions that scale like $\gs^{-6}$, i.e., $\scp=6$. Dimensions in which there are no branes have been eliminated from the table.}}
\end{table}

\begin{table}
\centering
\begin{tabular}{c||cccccccccc}
\hline\hline&&&&&&&&&&\\[-4mm]
$D\backslash p$ & $0$ & $1$ & $2$ & $3$ & $4$ & $5$ & $6$ & $7$ & $8$ & $9$\\
\hline\hline\\[-4mm]
$4$ &  &  &  & 32 &  &&&&\\
\hline&&&&&&&&&&\\[-4mm]
$3$ &  & 64 & 2688 &&&&&&& \\
\hline\hline

\end{tabular}
{\caption{\label{tab:scp7}\sl Supersymmetric $p$-branes in $D$ dimensions that scale like $\gs^{-7}$, i.e., $\scp=7$. Dimensions in which there are no branes have been eliminated from the table.}}
\end{table}

\begin{table}
\centering
$D=3$:\quad\quad
\begin{tabular}{c||cccccccccc}
\hline\hline&&&&&&&&&&\\[-4mm]
$\scp\backslash p$ & $0$ & $1$ & $2$ & $3$ & $4$ & $5$ & $6$ & $7$ & $8$ & $9$\\
\hline\hline\\[-4mm]
$8$ &   & 1 & 2304 & & &&&&\\
\hline&&&&&&&&&&\\[-4mm]
$9$ &   &  & 1344 &  &&&&&\\
\hline&&&&&&&&&&\\[-4mm]
$10$ &   &  & 560 &  &&&&&\\
\hline&&&&&&&&&&\\[-4mm]
$11$ &   & &64&&&&&&& \\
\hline\hline

\end{tabular}
{\caption{\label{tab:scpn}\sl Supersymmetric $p$-branes in $3$ dimensions that scale like $\gs^{-\scp}$ for $\scp=8,9,10,11$. This table is different from the preceding ones as only $D=3$ appears and we have therefore chosen a more compact presentation.}}
\end{table}

With these conventions the full list of mixed symmetry tensors in $D=11$ necessary for the counting in table~\ref{tab:SB} is 
\begin{subequations}
\label{eq:MOrigin}
\begin{align}
&T_{(10,1)}, T_{(3)}, T_{(6)}, T_{(8,1)},&\\
&T_{(9,3)}, T_{(9,6)}, T_{(9,8,1)},&\\
& T_{(10,1,1)},  T_{(10,4,1)},  T_{(10,6,2)},  T_{(10,7,4)}, T_{(10,7,7)}, T_{(10,8,2,1)}, T_{(10,8,5,1)}, T_{(10,8,7,2)},&\nn\\
&\quad\quad T_{(10,8,8,4)}, T_{(10,8,8,7)},& \\
&T_{(11,4,3)}, T_{(11,5,1,1)}, T_{(11,6,3,1)}, T_{(11,6,6,1)},T_{(11,7,4,2)}, T_{(11,7,6,3)}, T_{(11,7,7,5)},T_{(11,8,3,1,1)}, &\nn\\
&\quad\quad T_{(11,8,4,4)}, T_{(11,8,5,2,1)}, T_{(11,8,6,4,1)}, T_{(11,8,7,2,2)}, T_{(11,8,7,5,2)},T_{(11,8,7,7,3)}, &\nn\\
&\quad\quad T_{(11,8,8,1,1,1)}, T_{(11,8,8,4,1,1)}, T_{(8,8,5,4)}, T_{(11,8,8,6,2,1)}, T_{(11,8,8,7,4,1)}, T_{(11,8,8,7,7,1)},&\nn\\
&\quad\quad T_{(11,8,8,8,2,2)}, T_{(11,8,8,8,5,2)}, T_{(11,8,8,8,7,3)}, T_{(11,8,8,8,8,5)}, T_{(11,8,8,8,8,8)}.
\end{align}
\end{subequations}
We have grouped the generators into four classes in a specific way that arises from the form of their extremal (lowest) root vectors. The first line belongs to $E_8$, the second line to $E_9$, the third line to $E_{10}$ and the last line to $E_{11}$. Solutions corresponding to the first two lines with co-dimension at most two can be written in terms of  $D=11$ supergravity by using the techniques of~\cite{LozanoTellechea:2000mc,Englert:2007qb}. The last two lines are associated with deformations of supergravity~\cite{Riccioni:2007au,Bergshoeff:2007qi,Riccioni:2006az}.

The real root components of these mixed symmetry tensors are always obtained when the maximal number of indices is identical. This can be seen by considering the action of the lowering operators of $GL(11,\reals)$ on the tensors. The lowest root component of the multiplet is the one where all indices take their maximum value. For example, if we write the tensor $T_{(10,1,1)}$ explicitly with indices as 
\begin{align}
\label{eq:Romansmass}
T_{(10,1,1)}\,:\, T_{M_1\ldots M_{10},N,P},
\end{align}
then the lowest root vector component is ($M,N,P=1,\ldots,11$)
\begin{align}
T_{2\,\ldots 11,11,11}.
\end{align}
We see that this tensor, reduced along the direction $11$ will give rise to a pure nine-form. In fact, it is the nine-form that the mass of massive type IIA couples to~\cite{Bergshoeff:1996ui,Henneaux:2008nr}. We note that the highest $GL(11,\reals)$ level needed in $E_{11}$ to accommodate all the tensors of (\ref{eq:MOrigin}) is $\ell=17$. 

One can also analyse the T-duality representations that occur for the different values of $\scp$. For $\scp\leq 3$, these were already given in~\cite{Bergshoeff:2011zk,Bergshoeff:2011ee}. For $\scp=4$ we find that for $D\leq 7$ the following representations of $SO(10-D,10-D;\reals)$ arise\footnote{If $D=7$ the hook tableau does not exist and the representation is absent. Similarly, in the table for $\scp=5$.}
\begin{center}
\begin{tabular}{c|c}
$(D-2)$-brane & $(7-D)$-form \\\hline
$(D-1)$-brane & self-dual $(10-D)$-form and $(7-D,1)$ hook
\end{tabular}
\end{center}
In addition, there is a singlet $(D-3)$-brane in $D=4$.

For $\scp=5$ one finds the following representations of $SO(10-D,10-D;\reals)$ in $D\leq 7$.
\begin{center}
\begin{tabular}{c|c}
$(D-2)$-brane & $(4-D)$ tensor spinor\\\hline
$(D-1)$-brane & $(6-D)$ tensor spinor and $(4-D,1)$ hook spinor
\end{tabular}
\end{center}

For $\scp>5$ one can also easily determine the T-duality representations but since there are very few cases that appear, we do not list them here. For even $\scp$ one obtains tensors and for odd $\scp$ (general) tensor spinors.
To recover the correct counting from the T-duality representations one again has to restrict to real roots; alternatively, the correct counting is given by the size of the $SO(10-D,10-D;\reals)$ Weyl orbit of the lowest weight.

\section{Co-dimension two supergravity solutions}
\label{sec:sols}

The IIA/IIB solutions that are required as additional sources for the supersymmetric branes discussed in section~\ref{sec:e11sec} can be constructed from (discrete) U-duality. In the language of $E_{11}$ introduced above this corresponds to Weyl transformations. In this section we exemplarily construct some such solutions of co-dimension two. 

\subsection{$D=7$ and $1/\gs^2$}

As the first example we take 4-branes in $D=7$ maximal supergravity with tension $1/\gs^2$ in string frame. There are two transverse directions that we take to be in directions $1$ and $2$. The counting predicts a total of $12$ such objects. They have the following ten-dimensional origin 
\begin{itemize}
\item Three singly wrapped NS$5$-branes by choosing one of the world-volume directions to lie along one of the three compact directions.
\item Six wrapped Kaluza--Klein 6-monopoles. The NUT has to be in one of the three compact directions, and another world-volume direction is along one of the other two compact directions, leading to $3\times 2$ possibilities. 
\item Three `non-standard' branes. These are the ones that we will discuss in more detail now.
\end{itemize}

Considered as real roots of $E_{11}$ the three `non-standard' branes correspond to 
\begin{subequations}
\begin{align}
\beta_1 &= \alpha_3 + 2\alpha_4 + 3\alpha_5 + 4\alpha_6 + 5\alpha_7 
   + 6\alpha_8 + 4\alpha_9 + 2\alpha_{10} + 4\alpha_{11},&\\
\beta_2 &= \alpha_3 + 2\alpha_4 + 3\alpha_5 + 4\alpha_6 + 5\alpha_7 
   + 7\alpha_8 + 4\alpha_9 + 2\alpha_{10} + 4\alpha_{11},&\\
   \label{eq:b3}
\beta_3 &= \alpha_3 + 2\alpha_4 + 3\alpha_5 + 4\alpha_6 + 5\alpha_7 
   + 7\alpha_8 + 5\alpha_9 + 2\alpha_{10} + 4\alpha_{11}.&
\end{align}
\end{subequations}
{}From a IIA perspective they are all at level $(2,4)$ since the coefficients of $\alpha_{10}$ and $\alpha_{11}$ are $2$ and $4$, respectively. They all belong to the $GL(10,\reals)$ representation with an $(8,2)$ Young tableau. Considering the affine null root $\delta$ that has level $(2,3)$ in the IIA decomposition, we expect that all solutions are more complicated versions of the fundamental string that sits at level $(0,1)=(2,4)-(2,3)$ in the level decomposition. Additionally, we expect from the form of the roots that this string is extended along the directions $9$ and $10$.

Let us construct a solution corresponding to $\beta_1$. The fastest way of doing this is by realising $\beta_1$ as a short Weyl transformation acting on a known solution. We choose
\begin{align}
\label{lev4}
\beta_1 = w_8 w_{11} w_8 (\beta_0)
\end{align}
for $\beta_0 = \delta-\alpha_8$. The BPS seed solution corresponding to $\beta_0$ is a (smeared) KK6-monopole that is aligned as follows in IIA:
\begin{center}
\begin{tabular}{c|c|c|c|c|c|c|c|c|c}
1&2&$\hat{3}$&4&5&6&7&8&9&10\\\hline
&&$\times$&$\times$&$\times$&$\times$&$\times$&&N&$\times$
\end{tabular}
\end{center}
The hat on the direction $3$ indicates that we choose this direction to be the time direction, `N' indicates the NUT. The IIA metric of this solution is given in string frame by
\begin{align}
ds_{\beta_0}^2 &= H \lp \lp dx^1\rp^2+\lp dx^2\rp^2+\lp dx^8\rp^2 \rp    +H^{-1}\lp dx^9 -B dx^8\rp^2&\nn\\
&\quad  - \lp dx^3\rp^2 +\lp dx^4\rp^2 +\lp dx^5\rp^2 +\lp dx^6\rp^2 +\lp dx^7\rp^2 +\lp dx^{10}\rp^2 .&
\end{align}
All other fields vanish. The equations of motion are satisfied if $H$ is harmonic and $B$ is its conjugate harmonic\footnote{We use $\epsilon_{12}=+1$ and a sign choice was made in writing (\ref{eq:HB}). The equations are also solved for the opposite choice.}
\begin{align}
\label{eq:HB}
\partial_i H = -\sum_{j=1,2} \epsilon_{ij}\partial_j B.
\end{align}
We furthermore assume $H$ to be positive. 

The Weyl transformation $w_8$ corresponds to an interchange of direction $8$ and $9$; the Weyl transformation $w_{11}$ is a double T-duality in directions $9$ and $10$ with subsequent interchange of the two directions~\cite{Obers:1998fb}. Carrying out the transformation dictated by (\ref{lev4}) using the Buscher rules one obtains the following string frame configuration\footnote{This is essentially the same calculation as in~\cite{Englert:2007qb}.}
\begin{subequations}
\label{eq:gs2sol}
\begin{align}
ds_{\beta_1}^2 &= \tilde{H}  \lp \lp dx^9 \rp^2 +    \lp dx^{10} \rp^2 \rp+ H \lp \lp dx^1 \rp^2 +    \lp dx^2 \rp^2 \rp &\nn\\
&\quad  - \lp dx^3\rp^2 +\lp dx^4\rp^2 +\lp dx^5\rp^2 +\lp dx^6\rp^2 +\lp dx^7\rp^2 +\lp dx^{10}\rp^2 ,&\\
B_{9\,10} &= \tilde{B},&\\
\phi &= \frac12\log \tilde{H}.&
\end{align}
\end{subequations}
All other field components vanish. The functions $\tilde{H}$ and $\tilde{B}$ are defined by
\begin{align}
\label{eq:tHtBdef}
\tilde{H} = \frac{H}{H^2+B^2},\quad\quad \tilde{B} = \frac{B}{H^2+B^2}
\end{align}
and are also conjugate
\begin{align}
\label{eq:tHtB}
\partial_i \tilde{H} = \sum_{j=1,2} \epsilon_{ij}\partial_j \tilde{B}.
\end{align}
We see that our expectations regarding the configuration are borne out. The similarity to the fundamental string (\ref{eq:F1sol}) can be seen after rescaling to Einstein frame.

This type IIA solution is best interpreted as a solution of $D=7$ supergravity where one has reduced along the isometry directions $8$, $9$ and $10$. In this case it couples not to an $(8,2)$ mixed Young tableaux but to a pure $5$-form. The tension of this object in string frame can be shown to be $1/\gs^2$ by standard U-duality techniques and we note that this solution has already appeared in~\cite{LozanoTellechea:2000mc}.

\subsection{$D=7$ and $\gs^{-3}$}

Table~\ref{tab:scp3} predicts four $4$-branes in $D=7$ with tension proportional to $\gs^{-3}$. One of them is obtained by wrapping the S-dual of the D$7$ brane of type IIB, the origin of the other three branes is not immediate from the standard string theory branes. They could be obtained by $SL(5,\ints)$ U-duality from the wrapped S-dual in $D=7$. We instead construct them directly in type IIA by considering the corresponding $E_{11}$ roots. 

The four roots are
\begin{subequations}
\begin{align}
\beta_1 &= \alpha_3+2\alpha_4+3\alpha_5+4\alpha_6+5\alpha_7+6\alpha_8+4\alpha_9+3\alpha_{10}+3\alpha_{11},&\\
\beta_2 &= \alpha_3+2\alpha_4+3\alpha_5+4\alpha_6+5\alpha_7+6\alpha_8+5\alpha_9+3\alpha_{10}+3\alpha_{11},&\\
\beta_3 &= \alpha_3+2\alpha_4+3\alpha_5+4\alpha_6+5\alpha_7+7\alpha_8+5\alpha_9+3\alpha_{10}+3\alpha_{11},&\\
\label{eq:b4}
\beta_4 &= \alpha_3+2\alpha_4+3\alpha_5+4\alpha_6+5\alpha_7+7\alpha_8+5\alpha_9+3\alpha_{10}+4\alpha_{11}&
\end{align}
\end{subequations}
{}From the form of the roots, we can guess their type IIB intepretation. The first root is the one corresponding to the S-dual of the type IIB $7$-brane; the other three resemble more complicated versions of the D$1$-branes. Similarly, we can give their origin in $D=11$. Here, the first three arise from wrapping the KK7-monopole in specific ways, whereas $\beta_4$ is a more complicated $3$-brane.

In type IIA theory, the first three are similar to D$0$-particles whereas the last one is a kind of D$2$-brane. Let us construct the solution for $\beta_3$ and $\beta_4$ explicitly. For $\beta_3$ we do this by reduction of a suitable (smeared) KK7-monopole of $D=11$. From the root vector $\beta_3$ one deduces that the KK7-monopole in $D=11$ has to be
\begin{align}
ds_{11}^2 &= H \lp \lp dx^1 \rp^2 + \lp dx^2 \rp^2 +\lp dx^{11} \rp^2\rp  - \lp dx^3 \rp &\nn\\
&\quad\quad +\sum_{k=4,5,6,7,9,10} \lp dx^k\rp^2 + H^{-1} \lp dx^{8} - B dx^{11}\rp^2.&
\end{align}
The NUT is in the direction $8$ and time is in direction $3$. The functions $B$ and $H$ depend on $x^1$ and $x^2$ and satisfy (\ref{eq:HB}). This solution has to be reduced along the M-theory direction $x^{11}$. The result in IIA Einstein frame is
\begin{subequations}
\begin{align}
ds_{\beta_3}^2 &= \tilde{H}^{-1/8}H \lp \lp dx^1\rp^2+\lp dx^2\rp\rp^2 -\tilde{H}^{-1/8} \lp dx^3\rp^2 &\nn\\
&\quad\quad+\tilde{H}^{-1/8}\sum_{k=4,5,6,7,9,10} \lp dx^k\rp^2 +\tilde{H}^{7/8} \lp dx^{10}\rp^2,&\\
\phi &= -\frac34 \log \tilde{H},&\\
A_{8} &=- \tilde{B},&
\end{align}
\end{subequations}
where $\tilde{H}$ and $\tilde{B}$ are as in~(\ref{eq:tHtBdef}).
As expected, the solutions couples to the $1$-form potential and is hence similar to a D$0$-brane.

Since $\beta_4=w_{11}(\beta_3)$ we can construct the solution for $\beta_4$ by performing a double T-duality in directions $9$ and $10$ and subsequent exchange of directions $9$ and $10$. We need to use the Buscher rules extended to the RR sector since the IIA vector potential is non-vanishing. These have been studied for example in~\cite{Bergshoeff:1996ui}. A version sufficient for our purposes is derived in (\ref{IIAIIBEF}) in the appendix.
The resulting solution of type IIA in Einstein frame is
\begin{subequations}\label{eq:gs3sol}
\begin{align}
ds_{\beta_4}^2 &= \tilde{H}^{-3/8}H \lp \lp dx^1\rp^2+\lp dx^2\rp\rp^2 -\tilde{H}^{-3/8} \lp dx^3\rp^2 &\nn\\
&\quad\quad+\tilde{H}^{-3/8}\sum_{k=4,5,6,7,9,10} \lp dx^k\rp^2 
+\tilde{H}^{5/8} \lp\lp dx^{10}\rp^2 + \lp dx^9\rp^2+\lp dx^{10}\rp^2\rp,&\\
\phi &= -\frac14 \log \tilde{H},&\\
A_{8\,9\,10} &=- \tilde{B}.&
\end{align}
\end{subequations}
The solution couples to the $3$-form potential of type IIA as expected and also displays a metric characteristic of a D$2$-type object. This solution has appeared in slightly different form in~\cite{LozanoTellechea:2000mc}, where it was also shown that its tension in string frame scales like $1/\gs^3$.

\subsection{$D=4$ and $\gs^{-4}$}

As one example for string frame tension $1/\gs^{4}$ we study a co-dimension two brane in $D=4$. According to table~\ref{tab:scp4} there is a single such object. Its root vector is
\begin{align}
\label{eq:bscp4}
\beta = \alpha_3+ 2\alpha_4+ 4\alpha_5+ 6\alpha_6+ 8\alpha_7+ 10\alpha_8+ 7\alpha_9+ 4\alpha_{10}+ 5\alpha_{11}.
\end{align}
The origin of this solution in type IIA and type IIB is that of a more complicated NS$5$-brane, whereas in $D=11$ it is related to the M$5$-brane. We construct the solution in type IIA by starting from a known solution and performing an appropriate series of duality transformations.

One way of arriving at $\beta$ is by starting from 
\begin{align}
\beta_0 = \alpha_3 + 2\alpha_4 + 3\alpha_5 + 5\alpha_6 + 7\alpha_7 
   + 9\alpha_8 + 6\alpha_9 + 3\alpha_{10} + 4\alpha_{11}
\end{align}
(which agrees with $w_8w_9w_7w_8w_6w_7(\beta_4$) with $\beta_4$ of (\ref{eq:b4})) and then considering the following chain of $E_{11}$ Weyl transformations
\begin{align}\label{eq:gs4chain}
\beta = w_{10} w_9w_8w_7w_6w_5w_{11}(\beta_0).
\end{align}
The only transformations that do not correspond simply to a permutation of space-like directions in type IIA are the ones in with indices $10$ and $11$; we will deal with them by lifting the solution corresponding to $\beta_0$ to $D=11$ where the reflection in $10$ is also a simple permutation.\footnote{Note that is the final reflection in $10$ that changes the scaling of the tension with $\gs$.} The reflection in $11$ is a, as before, a double T-duality in type IIA in directions $9$ and $10$ with interchange of the two directions.

The type IIA Einstein frame solution corresponding to $\beta_0$ (obtained as a permutation variant of (\ref{eq:gs3sol})) is
\begin{subequations}
\begin{align}
ds^2_{\beta_0} &= H\tilde{H}^{-3/8} \lp \lp dx^1\rp^2+\lp dx^2\rp^2\rp - \tilde{H}^{-3/8}\lp dx^3\rp^2&\nn\\
&\quad +\tilde{H}^{-3/8}\sum_{k=4,5,9,10} \lp dx^k\rp^2 +\tilde{H}^{5/8}\lp \lp dx^6\rp^2 +\lp dx^7\rp^2+\lp dx^8\rp^2\rp,&\\
\phi &=-\frac14\log \tilde{H},&\\
A_{6\,7\,8} &=-\tilde{B}.&
\end{align}
\end{subequations}
The $w_{11}$ reflection gives the metric IIA Einstein frame solution
\begin{subequations}
\begin{align}
ds^2_{w_{11}(\beta_0)} &= H\tilde{H}^{-5/8} \lp \lp dx^1\rp^2+\lp dx^2\rp^2\rp  +\tilde{H}^{3/8}\sum_{k-6,7,8,9,10} \lp dx^k\rp^2
&\nn\\
&\quad +\tilde{H}^{-5/8}\lp-\lp dx^3\rp^2 +\lp dx^4\rp^2+\lp dx^5\rp^2\rp ,&\\
\phi &=\frac14\log \tilde{H},&\\
A_{3\,4\,5} &=-\frac1{\tilde{H}}.&
\end{align}
\end{subequations}
Performing now the remaining transformations of (\ref{eq:gs4chain}) one arrives finally at the solution for $\beta$ in type IIA Einstein frame
\begin{subequations}
\label{eq:gs4sol}
\begin{align}
ds^2_{\beta} &= H\tilde{H}^{-3/4} \lp \lp dx^1\rp^2+\lp dx^2\rp^2\rp - \tilde{H}^{-3/4}\lp dx^3\rp^2+\tilde{H}^{-3/4}\lp dx^4\rp^2&\nn\\
&\quad +\tilde{H}^{1/4}\sum_{k=5}^{10} \lp dx^k\rp^2 ,&\\
\phi &=-\frac12\log \tilde{H},&\\
B_{3\,4} &=-\frac1{\tilde{H}}.&
\end{align}
\end{subequations}
We see that it indeed couples to the NS--NS $2$-form as expected and has a metric similar to that of an NS $5$-brane. By the usual U-duality arguments its tension in string frame scales like $1/\gs^4$. This solution has been already listed in~\cite{LozanoTellechea:2000mc}.

\subsection{$D=3$ and $\gs^{-4}$}

As a final example we consider another $1/\gs^4$ brane, this time a $0$-brane in $D=3$. Out of the $14$ such objects in table~\ref{tab:scp4}, $7$ correspond more complicated NS $5$-branes of type IIA (obtained by reduction of solutions of the type studied just now) whereas there are $7$ purely gravitational solutions that we now aim to construct. They correspond to tensors of shape $(8,7,1)$ in type IIA and are hence generalised Kaluza--Klein monopoles.

One $E_{11}$ root that corresponds to such an object is
\begin{align}
\beta_{\rm G} = \alpha_3+ 3\alpha_4+ 5\alpha_5+7\alpha_6+ 9\alpha_7+ 11\alpha_8+ 7\alpha_9+ 4\alpha_{10}+ 6\alpha_{11}.
\end{align}
It can be reached from $\beta$ of (\ref{eq:bscp4}) by the following chain of Weyl transformations:
\begin{align}
\beta_{\rm G} = w_{11}w_8w_7w_6w_5w_4(\beta).
\end{align}
All of these operations we can carry out easily as permutations and (double) T-duality transformation in type IIA. What one arrives at is the following Einstein frame metric\footnote{Note the absence of $-\lp dx^3\rp^2$ in the metric which stems from the cancellation of terms when applying the T-duality rules (\ref{IIAIIBEF}).}
\begin{align}
ds^2_{\beta_{\rm G}} &= H\tilde{H}^{-1} \lp \lp dx^1\rp^2+\lp dx^2\rp^2\rp +\sum_{k=4}^9 \lp dx^k\rp^2
 +\tilde{H} \lp dx^{10} \rp^2 +2 dx^3 dx^{10}.&
\end{align}
All other fields are zero as this solution is purely gravitational. As we only performed T-dualities on the solution (\ref{eq:gs4sol}), the solution also has a string frame tension $1/\gs^4$. This solution was given before in~\cite{LozanoTellechea:2000mc}.

\subsection{Relation to Geroch group}

The proper language for analysing solutions that depend on two (or less) transverse direction is that of the Geroch group~\cite{Geroch:1970nt,Geroch:1972yt,Breitenlohner:1986um}. The Geroch group is the infinite-dimensional solution generating group of supergravity solutions that have $D-2$ commuting Killing vectors. It is an affine symmetry group~\cite{Julia:1982gx,Breitenlohner:1986um}; in the present case, it is affine $E_8$, i.e., $E_9$~\cite{Julia:1982gx,Nicolai:1987kz}. All the solutions relevant here are in fact described by an affine $SL(2,\reals)$ subgroup of $E_9$.

A convenient quantity for parametrising such solutions is the complex Ernst potential that we here define as~\cite{Englert:2007qb}\footnote{This complex function of two real fields is sufficient as long as one stays in an $SL(2,\reals)$ subgroup as we are doing for the elementary BPS solutions.}
\begin{align}
\label{eq:Ernst}
\cE = B+iH.
\end{align}
The conjugate harmonic equation (\ref{eq:HB}) then becomes the Cauchy--Riemann condition for $\cE$ to be holomorphic in $\xi=x^1+i x^2$. The transformed Ernst potential $\tilde{\cE}=\tilde{B}+i\tilde{H}$ then is anti-holomorphic in $\xi$ according to (\ref{eq:tHtB}). Furthermore, the two Ernst potentials are related by
\begin{align}
\tilde{\cE} = \frac{1}{\bar{\cE}}.
\end{align}
Thinking of $\cE$ as parametrising the upper half plane, this corresponds to the action of the matrix
\begin{align}
\left(\begin{array}{cc}1&0\\0&-1\end{array}\right)
\in PGL(2,\ints)
\end{align}
on it, where a negative determinant amounts to an additional complex conjugation of $\cE$. 
 
Any holomorphic Ernst potential $\cE$ corresponds to a BPS solution~\cite{Englert:2007qb} and hence any holomorphic transformation of $\cE$ will give a new BPS solution. Infinite towers of supersymmetric solutions were constructed using this approach in~\cite{Englert:2007qb}. However, the way of constructing these solutions required going to higher and higher level in the affine group. The mixed symmetry tensors in the affine group embedded in $E_{11}$ are of the form~\cite{Damour:2002cu,Kleinschmidt:2005bq,Riccioni:2006az}
\begin{align}
T_{(9,9,\ldots,9,3)},T_{(9,9,\ldots,9,6)}, T_{(9,9,\ldots,9,8,1)},
\end{align}
where we have used the $GL(11,\reals)\subset E_{11}$ decomposition that is appropriate for $D=11$ supergravity. The number of repetitions of columns of length $9$ is equal to the affine level $\ell_{\rm aff}$. We see that only up to affine level equal to $\ell_{\rm aff}=1$ it is possible to obtain forms in $D\geq 3$ dimensions. The other supersymmetric solutions of co-dimension two in the infinite towers of~\cite{Englert:2007qb} are brane solutions only in $D=2$ and therefore do not contribute to table~\ref{tab:SB}.\footnote{An additional complication beyond $\ell_{\rm aff}=1$ is that one also has to perform T-dualities in a time-like direction. Such T-dualities can change the signature of the theory~\cite{Hull:1998br,Hull:1998ym,Keurentjes:2004bv} and some of the infinitely many solutions of~\cite{Englert:2007qb} are therefore solutions of theories with different signature.}

\section{Co-dimension less than two solutions?}

In this section, we address the issue of constructing supergravity solutions with co-dimension equal to one or zero. We will see by means of an example that the construction of these objects is not always possible without also considering deformations of maximal supergravity.

The example we consider are space-filling branes in $D=8$ with string frame tensions scaling like $\gs^{-3}$.
Table~\ref{tab:scp3} predicts two such objects, the corresponding real roots are
\begin{subequations}
\label{d8s3}
\begin{align}
\label{d8s3a}
\beta_1 &= \alpha_1+ 2\alpha_2+ 3\alpha_3+ 4\alpha_4+ 5\alpha_5+ 6\alpha_6+7\alpha_7+ 8\alpha_8+ 4\alpha_9+ 3\alpha_{10}+ 4\alpha_{11},&\\
\beta_2 &= \alpha_1+ 2\alpha_2+ 3\alpha_3+ 4\alpha_4+ 5\alpha_5+ 6\alpha_6+7\alpha_7+ 8\alpha_8+ 7\alpha_9+ 3\alpha_{10}+ 4\alpha_{11}.&
\end{align}
\end{subequations}
They only differ in the $\alpha_9$ coefficient. Viewed from $D=11$ they are part of the level four tensor with Young tableau $(10,1,1)$, i.e., the Romans mass of (\ref{eq:Romansmass}). This representation does not correspond to any of the gradient representations of~\cite{Damour:2002cu} or to the the dual fields of~\cite{Riccioni:2006az}. But it is known to be the one that gives rise to the Romans mass~\cite{Romans:1985tz} upon dimensional reduction to $D=10$~\cite{Henneaux:2008nr}. Viewed from a type IIA perspective the roots (\ref{d8s3}) belong to a similar $(9,1,1)$ representation of $GL(10,\reals)$.

However, when viewed as coming from type IIB, the two roots are very different due to the different $\alpha_9$ coefficient. The root $\beta_1$ belongs to a pure eight-form potential in IIB, whereas $\beta_2$ belongs to a tableau of shape $(10,2,2)$ at level seven. This is indeed what is to be expected since we know that there is a space-filling $\gs^{-3}$ brane in type IIB already in $D=10$, namely the S-dual of the D$7$-brane. Its dimensional reduction to $D=8$ will give a similar brane in $D=8$ and it is the one associated with $\beta_1$. $\beta_2$, however, does not have such a simple explanation and it represents the new generalised KK-monopole required for making the counting correct. 

Attempting to construct the local solution corresponding to $\beta_2$ we consider a sequence of Weyl transformations that relate $\beta_2$ to known solutions. The following relation using a fundamental Weyl reflection  gives the solution
\begin{align}
\beta_2 = w_9(\beta_1).
\end{align}
Since we know what IIB solution $\beta_1$ corresponds to, and $w_9$ is represented as the exchange of directions $9$ and $10$ in IIA language, all we would need to do is to take the $\beta_1$ type IIB solution (S-dual of D$7$), translate it into IIA variables and then exchange directions $9$ and $10$. 

When trying to do this in practice one encounters the problem that the necessary IIA version of the D$7$ brane (or its S-dual) is the D$8$-brane (or its `S-dual'). Whereas the D$7$-brane is a bona fide solution of type IIB supergravity with two transverse directions on which all quantities depend holomorphically, the D$8$ brane is not a solution of the usual type IIA: One needs to consider the massive deformation of Romans to accommodate the D$8$-brane~\cite{Polchinski:1995mt}. That the D$8$-brane does not arise in type IIA can be seen as follows. In order to T-dualise one needs an isometry direction in the transverse space. However, since all everything depends holomorphically on the transverse space, turning one of the two directions isometric will render all functions constant and then the IIB solution is diffeomorphic to flat space. This can be circumvented in massive IIA theory~\cite{Bergshoeff:1996ui}. Therefore, we see that in order to describe all the supersymmetric branes of table~\ref{tab:SB}, we need to include all possible deformations (massive/gaugings) of maximal supergravity as well. We leave the construction of these solutions in gauged maximal supergravity or massive supergravity to future work.

\section{Conclusions}

In this paper we have determined the number of supersymmetric $p$-branes associated with a $(p+1)$-form potential in $3\leq D\leq 10$ dimensional maximal supergravity. The rule for finding the correct number rested on identifying supersymmetric branes as associated with real roots of $E_{11}$ as imaginary roots do not correspond to supersymmetric configurations~\cite{Houart:2011sk}. The numbers we have obtained in this way coincide perfectly with the counting done by completely different means for the cases that were analysed in~\cite{Bergshoeff:2011mh,Bergshoeff:2011ee}. Our results are complete and give predictions for supersymmetric branes that become very heavy at weak string coupling. For all supersymmetric branes we have determined their origin in ten-dimensional IIA and IIB theory and M-theory in terms of mixed symmetry tensors. 

All the new branes that are found in this way have co-dimension at most two. For the co-dimension two case we have explained how to identify the local supergravity solutions and have derived them by analysing Weyl orbits in $E_{11}$. The solutions thus obtained had already been derived differently in the literature in~\cite{LozanoTellechea:2000mc}. We have not addressed the issue of constructing finite energy solutions by suitable superpositions, i.e., choices of the Ernst potential in~(\ref{eq:Ernst}) but suspect that this will go very much along the lines of~\cite{Greene:1989ya}. We also see how to associate the co-dimension two solutions to mixed symmetry tensors of maximal supergravity. This we did for the example of type IIA and have exhausted the list of affine generators in (\ref{eq:AOrigin}). Clearly, the same can be done for type IIB and $D=11$ supergravity. 

For solutions with co-dimension at most one, the treatment given here in terms of massless or ungauged supergravity is generally not sufficient and the formalism will have to be extended to construct the new solutions. The role of all the new solutions for microstate counting remains to be clarified.

\vspace{3mm}
{\bf Acknowledgements}

\noindent The author would like to thank E.~Bergshoeff for valuable correspondence and S.~Theisen for useful discussions.

\vspace{3mm}
{\bf Note added:}
 After this paper was finished, the preprint~\cite{Bergshoeffnew} appeared that extends the Wess--Zumino analysis of~\cite{Bergshoeff:2010xc,Bergshoeff:2011zk,Bergshoeff:2011mh,Bergshoeff:2011ee} for $D\geq 6$. Where comparable those results are also in agreement with ours. Similarly, the paper~\cite{BOR} appeared after first release of this paper and contains some overlap with the results discussed in section~\ref{sec:sols}.

\appendix

\section{Type IIA and IIB origin of all solutions}
\label{app:IIAIIBorigin}

In this appendix, we give the complete list of all tensors in type IIA and type IIB necessary for accommodating all supersymmetric solutions summarised in table~\ref{tab:SB}.

{\allowdisplaybreaks
The type IIA mixed tensor fields necessary for accommodating all the solutions are:
\begin{subequations}
\label{eq:AOrigin}
\begin{align}
& T_{(9,1)},T_{(1)}, T_{(2)}, T_{(3)}, T_{(5)}, T_{(6)}, T_{(7)}, T_{(7,1)}, &\\
& T_{(8,1)}, T_{(8,2)},T_{(8,3)}, T_{(8,5)}, T_{(8,6)}, T_{(8,7)}, T_{(8,7,1)},&\\
&T_{(9)}, T_{(9,1,1)},T_{(9,3)}, T_{(9,3,1)}, T_{(9,3,3)}, T_{(9,4,1)}, T_{(9,5,1)}, T_{(9,5,2)},T_{(9,6)}, T_{(9,6,2)}, T_{(9,6,3)}, &\nn\\
&\quad\quad  T_{(9,6,4)},T_{(9,6,5,2)}, T_{(9,6,6)}, T_{(9,7,1)}, T_{(9,7,1,1)}, T_{(9,7,2,1)}, T_{(9,7,4)}, T_{(9,7,4,1)},  T_{(9,7,5,1)}, &\nn\\
&\quad\quad T_{(9,7,6,1)}, T_{(9,7,6,2)}, T_{(9,7,7)}, T_{(9,9,7,2)}, T_{(9,7,7,3)}, T_{(9,7,7,6)}, T_{(9,7,7)},&\\
&T_{(10,1,1)} , T_{(10,3,2)}, T_{(10,3,3)}, T_{(10,4)},T_{(10,4,1)},  T_{(10,4,1,1)}, T_{(10,4,3)}, T_{(10,5,1,1)},  T_{(10,5,2)},&\nn\\ 
&\quad\quad T_{(10,5,2,1)}, T_{(10,5,3,1)}, T_{(10,5,5)}, T_{(10,5,5,1)}, T_{(10,6,2)}, T_{(10,6,3,1)}, T_{(10,6,3,2)},&\nn\\
&\quad\quad T_{(10,6,4,2)},  T_{(10,6,5,2)}, T_{(10,6,5,3)}, T_{(10,6,6,1)}, T_{(10,6,6,3)}, T_{(10,6,6,4)}, T_{(10,6,6,5)},&\nn\\
&\quad\quad T_{(10,7,2)},T_{(10,7,2,1,1)}, T_{(10,7,3,1,1)}, T_{(10,7,3,3)},T_{(10,7,4)},T_{(10,7,4,1)}, T_{(10,7,4,1,1)}, &\nn\\
&\quad\quad T_{(10,7,4,2)},T_{(10,7,4,2,1)}, T_{(10,7,4,4)}, T_{(10,7,5,2,1)}, T_{(10,7,5,3)}, T_{(10,7,5,3,1)},&\nn\\
&\quad\quad T_{(10,7,5,4,1)}, T_{(10,7,6,1,1)}, T_{(10,7,6,2,2)}, T_{(10,7,6,3)}, T_{(10,7,6,4,1)}, T_{(10,7,6,4,2)},&\nn\\
&\quad\quad T_{(10,7,6,5,2)}, T_{(10,7,6,6,3)}, T_{(10,7,7)},  T_{(10,7,7,1,1,1)}, T_{(10,7,7,2,2)}, T_{(10,7,7,3)}, &\nn\\
&\quad\quad T_{(10,7,7,3,1,1)}, T_{(10,7,7,4,1,1)}, T_{(10,7,7,4,3)}, T_{(10,7,7,4,4)}, T_{(10,7,7,5)}, T_{(10,7,7,5,1)},&\nn\\
&\quad\quad T_{(10,7,7,5,1,1)}, T_{(10,7,7,5,2)}, T_{(10,7,7,5,2,1)}, T_{(10,7,7,5,4)}, T_{(10,7,7,6,2,1)}, T_{(10,7,7,6,3)},&\nn\\
&\quad\quad T_{(10,7,7,6,3,1)}, T_{(10,7,7,6,4,1)}, T_{(10,7,7,6,6)},  T_{(10,7,7,6,6,1)}, T_{(10,7,7,7,1,1)}, T_{(10,7,7,7,2,2)},&\nn\\
&\quad\quad  T_{(10,7,7,7,3)}, T_{(10,7,7,7,4,1)}, T_{(10,7,7,7,4,2)}, T_{(10,7,7,7,5,2)}, T_{(10,7,7,7,6,2)}, T_{(10,7,7,7,6,3)},&\nn\\
&\quad\quad T_{(10,7,7,7,7,1)}, T_{(10,7,7,7,7,3)}, T_{(10,7,7,7,7,4)}, T_{(10,7,7,7,7,5)}, T_{(10,7,7,7,7,7)}.&
\end{align}
\end{subequations}
As in (\ref{eq:MOrigin}), we have grouped them into four classes according to the origin of their lowest root vectors in $E_8$, $E_9$, $E_{10}$ or $E_{11}$.

The higher-dimensional of all the branes of table~\ref{tab:SB} can be found in type IIB by considering the following set of tensors:
\begin{subequations}
\label{eq:BOrigin}
\begin{align}
&T_{(9,1)}, T_{(0)}, T_{(2)}, T_{(4)}, T_{(6)}, T_{(7,1)}, &\\
& T_{(8)}, T_{(8)}'. T_{(8,2)}, T_{(8,4)}, T_{(8,6)}, T_{(8,7,1)},&\\
& T_{(9,2,1)}, T_{(9,3)}, T_{(9,4,1)}, T_{(9,5,2)}, T_{(9,6,1)}, T_{(9,6,3)}, T_{(9,6,5)}, T_{(9,7,1,1)}, T_{(9,7,3,1)},&\nn\\
&\quad\quad T_{(9,7,4)}, T_{(9,7,5,1)}, T_{(9,7,6,2)}, T_{(9,7,7,1)}, T_{(9,7,7,3)}, T_{(9,7,7,5)}, T_{(9,7,7,7)}&\\
&  T_{(10)}, T_{(10)}',T_{(10,2,2)}, T_{(10,4)}, T_{(10,4,1,1)}, T_{(10,4,2)}, T_{(10,4,4)},  T_{(10,5,2,1)}, T_{(10,5,4,1)},&\nn\\
&\quad\quad T_{(10,6,2)}, T_{(10,6,2,2)}, T_{(10,6,3,1)}, T_{(10,6,4,2)}, T_{(10,6,5,1)}, T_{(10,6,5,3)}, T_{(10,6,6)},&\nn\\
&\quad\quad  T_{(10,6,6,2)}, T_{(10,6,6,4)}, T_{(10,6,6,6)}, T_{(10,7,1,1,1)}, T_{(10,7,3,1,1)}, T_{(10,7,4,1)}, T_{(10,7,4,2,1)}, &\nn\\
&\quad\quad T_{(10,7,4,3)}, T_{(10,7,5,1,1)}, T_{(10,7,5,3,1)}, T_{(10,7,5,5,1)}, T_{(10,7,6,3)}, T_{(10,7,6,3,2)}, T_{(10,7,6,4,1)},&\nn\\
&\quad\quad T_{(10,7,6,5,2)}, T_{(10,7,6,6,3)}, T_{(10,7,7,1,1)}, T_{(10,7,7,2,1,1)}, T_{(10,7,7,3,3)}, T_{(10,7,7,4,1,1)},&\nn\\
&\quad\quad T_{(10,7,7,5,1)}, T_{(10,7,7,5,2,1)}, T_{(10,7,7,5,3)}, T_{(10,7,7,5,5)}, T_{(10,7,7,6,1,1)}, T_{(10,7,7,6,3,1)},&\nn\\
&\quad\quad T_{(10,7,7,6,5,1)}, T_{(10,7,7,7,3)}, T_{(10,7,7,7,3,2)}, T_{(10,7,7,7,4,1)}, T_{(10,7,7,7,5,2)}, T_{(10,7,7,7,6,3)},&\nn\\
&\quad\quad T_{(10,7,7,7,7)}, T_{(10,7,7,7,7,2)}, T_{(10,7,7,7,7,4)}, T_{(10,7,7,7,7,6)}.&
\end{align}
\end{subequations}}

\section{Conventions}
\label{app:conv}

We use the mostly plus convention for the metric.

\subsection{Type IIA and type IIB}

Our conventions for type IIA and type IIB supergravity (in a non-democratic formulation) are as follows. The NS--NS sector action in string frame is
\begin{align}
\label{NSstring}
S_{\rm NS-NS} = \int d^{10}x \sqrt{-g^{\rm s}} e^{-2\phi} \lp R^{\rm s} +4 (\partial \phi)^2 -\frac1{12} H_{(3)}^2\rp,
\end{align}
where we have eliminated Planck's constant and the pre-factor. Upon changing to the Einstein frame metric $g_{mn}^{\rm E} = e^{- \phi/2} g_{mn}^{\rm s}$, the action becomes
\begin{align}
S_{\rm NS-NS} = \int d^{10}x \sqrt{-g^{\rm E}} \lp R^{\rm E} -\frac12 (\partial \phi)^2 -\frac1{12}e^{-\phi} H_{(3)}^2\rp.
\end{align}
Here, $H_{(3)}^2= H_{mnp}H^{mnp}$.
The NS--NS sector is common to type IIA and type IIB, however, it will be useful to distinguish the fields in the two theories. We put hats on all type IIB quantities.

The string frame action for the R--R sector of IIA is given by
\begin{align}
\label{RA}
S_{\rm R-R}^{\rm (IIA)} &= \int d^{10}x \sqrt{-g^s} 
  \left(-\frac14 F_{(2)}^2 - \frac1{48} F_{(4)}^2\right).
\end{align}

For type IIB it is given by the pseudo-action (of course, to be supplemented by self-duality of the five-form)\footnote{In string frame, the $SL(2,\reals)$ symmetry of type IIB  is not manifest since the metric transforms. In Einstein frame, the kinetic terms for the two scalars combine as usual into $-\frac12\tau_2^{-2} \partial\tau\partial\bar{\tau}$, where $\tau=\hat{\chi}+ie^{-\hat{\phi}}$ and $\partial\hat{\chi}=\hat{F}_{(1)}$ and $SL(2,\reals)$ acts by fractional linear transformations on $\tau$. The modified two $3$-form field strengths $(\hat{F}_{(3)}+\hat{\chi} \hat{H}_{(3)},\hat{H}_{(3)})$ form an $SL(2,\reals)$ doublet. It is the modified field strengths that satisfy trivial Bianchi identities (and hence derive directly from $2$-form potentials that also transform as a linear doublet). The $5$-form potential $\hat{F}_{(5)}$ is invariant but satisfies a non-trivial Bianchi identity~\cite{Schwarz:1983wa}. All this also follows from the structure of the exceptional algebra~\cite{Schnakenburg:2001ya,Kleinschmidt:2004rg}.}
\begin{align}
\label{RB}
S_{\rm R-R}^{\rm (IIB)} &= \int d^{10}x \sqrt{-\hat{g}^s} 
  \left(-\frac12 \hat{F}_{(1)}^2 - \frac1{12} \hat{F}_{(3)}^2-\frac1{480} \hat{F}_{(5)}^2\right).
\end{align}
In both actions, $F_{(p+1)}$ denotes a field strength with $p+1$ anti-symmetric indices arising from a $p$-form potential; its square is constructed by full contraction with the metric.
We do not give the Chern--Simons terms as they will vanish for all solutions considered in this paper. Similarly, we will not give the `transgression' and Chern--Simons type terms that appear in the definition of the various field strengths.

In the presence of an isometry, the two theories become equivalent and one can map one set of variables to the other.\footnote{This is clear since there is only one maximal ungauged supergravity theory in $D=9$. Relations of the type below have already been derived before, see for instance~\cite{Bergshoeff:1995as}.} Assume that $z=10$ is the isometry direction. The map between the two reduced theories is most easily done in the Einstein frame. We write the reduction ansatz in terms of the zehnbein decomposed as a block diagonal matrix\footnote{We omit the Einstein frame superscript in order to keep the expressions more legible. Our reduction ansatz gives the Einstein frame in $D=9$.}
\begin{align}
e_m{}^a = \left(\begin{array}{cc}
e^{c \newphi} e_\mu{}^\alpha & e^{-7c\newphi} \cA_\mu\\
0 & e^{-7c \newphi}
\end{array}\right),\quad\quad
\text{where}\quad c=\frac1{4\sqrt{7}}.
\end{align}
The field strength of the Kaluza--Klein vector $\cA$ will be denoted by $\cF$ in order to distinguish it from the forms arising from the reduction of the various $p$-form fields. With this ansatz the NS--NS sector action (\ref{NSstring}) reduces to
\begin{align}\label{NSred}
S_{\rm NS-NS} &= \int d^9x  \sqrt{-g_{(9)}} 
\left(R_{(9)}-\frac12 (\partial\newphi)^2-\frac12(\partial\phi)^2\right.&\nn\\
&\quad\left. -\frac14e^{-\phi+\frac3{\sqrt{7}}\newphi} H_{(2)}^2-\frac14 e^{-\frac4{\sqrt{7}}\newphi} \cF_{(2)}^2-\frac1{12}e^{-\phi-\frac1{\sqrt{7}}\newphi}H_{(3)}^2
\right),
\end{align}
where we have set the volume of the isometry direction to one. Again, we do not give the full definitions of the various field strengths as they will not matter in our solutions. The reduced action is valid for both type IIA and IIB; one has to put hats on {\em all} variables for type IIB.

The type IIA R--R sector (\ref{RA}) reduces in Einstein frame to
\begin{align}
S_{\rm R-R}^{\rm (IIA)} &=\int d^9x \sqrt{-g_{(9)}} 
\left(-\frac12e^{\frac32\phi+\frac{\sqrt{7}}{2}\newphi}F_{(1)}^2-\frac14e^{\frac32\phi-\frac1{2\sqrt{7}}\newphi}F_{(2)}^2\right.&\nn\\
&\left.\quad\quad\quad\quad
-\frac1{12}e^{\frac12\phi+\frac5{2\sqrt{7}}\newphi}F_{(3)}^2-\frac1{48}e^{\frac12\phi-\frac{3}{2\sqrt{7}}\newphi}F_{(4)}^2
\right),&
\end{align}
whereas the type IIB R--R sector action (\ref{RB}) reduces to\footnote{Note that the pseudo-action is transformed into a proper action as there is no longer any problem with self-duality in nine dimensions. The five-form in nine dimensions was dualised into a four-form.}
\begin{align}
S_{\rm R-R}^{\rm (IIB)} &= \int d^9x \sqrt{-\hat{g}_{(9)}}
\left( -\frac12 e^{2\hat{\phi}}\hat{F}_{(1)}^2 -\frac14e^{\hat{\phi}+\frac3{\sqrt{7}}\hat{\newphi}}\hat{F}_{(2)}^2\right.&\nn\\
&\left.\quad\quad\quad\quad
-\frac1{12}e^{\hat{\phi}-\frac1{\sqrt{7}}\hat{\newphi}}\hat{F}_{(3)}^2-\frac1{48}e^{\frac2{\sqrt{7}}\hat{\newphi}}\hat{F}_{(4)}^2
\right).&
\end{align}
The map between the two theories (in Einstein frame) is now easily deduced to be\footnote{The relations for the field strengths can be directly integrated up to the corresponding potentials, ignoring Chern--Simons and transgression terms.}
\begin{subequations}
\label{IIAIIBEF}
\begin{align}
g_{(9)} &= \hat{g}_{(9)},& H_{(3)} &=\hat{H}_{(3)},&\\
F_{(1)} &= \hat{F}_{(1)},& F_{(4)} &= \hat{F}_{(4)},& \\
 F_{(2)} &=\hat{F}_{(2)},& F_{(3)} &=\hat{F}_{(3)},&\\
\cF_{(2)} &= \hat{H}_{(2)},& H_{(2)} &=\hat{\cF}_{(2)},&
\end{align}
together with the (orthogonal) transformation of dilatons
\begin{align}
\phi &= \frac34\hat{\phi}+\frac{\sqrt{7}}{4}\hat{\newphi},&
\newphi &= \frac{\sqrt{7}}{4}\hat{\phi}-\frac{3}{4}\hat{\newphi}.&
\end{align}
\end{subequations}

\subsection{(Some) known solutions}

The fundamental string extended in directions $9$ and $10$ and smeared such that the only remaining transverse directions are $1$ and $2$ is written in Einstein frame as~\cite{Dabholkar:1990yf}
\begin{subequations}
\label{eq:F1sol}
\begin{align}\label{F1metric}
ds^2_{\rm F1} &=  H^{-3/4}\lp \lp dx^9\rp^2-\lp dx^{10}\rp^2\rp +H^{1/4}\sum_{i=1,\ldots, 8} \lp dx^i\rp^2,&\\
\phi &=-\frac12\log H,&\\
B_{9\,10} &= 1/H,&
\end{align}
\end{subequations}
where $H$ is harmonic. As it only uses the NS--NS fields, it is a solution of type IIA and IIB identical in form. 

Interpreting it as a solution of type IIB we can perform an S-duality transformation. The result is the D$1$ string solution. Its metric is identical to (\ref{F1metric}) but the dilaton $\phi$ changes sign and $B_{9\,10}$ is replaced by the Ramond potential $A_{9\,10}$.\footnote{As $\chi=0$ for the fundamental string, the modified and unmodified field strengths are identical.}

The D$7$ brane of type IIB is given by the following expression in Einstein frame
\begin{subequations}
\label{D7solution}
\begin{align}\label{D7metric}
ds^2_{\rm D7} &=  \tau_2\lp \lp dx^9\rp^2+\lp dx^{10}\rp^2\rp -\lp dx^3\rp^2+\sum_{i=1,2,4,\ldots, 8} \lp dx^i\rp^2,&\\
\tau &= \chi+i e^{-\phi}&
\end{align}
\end{subequations}
for any holomorphic (or anti-holomorphic) $\tau=\tau(x^1+i x^2)$. We have aligned this solution such that the transverse space is in directions $9$ and $10$ and time is in direction $3$. In this way it corresponds to the $E_{11}$ root 
\begin{align}
\label{D7root}
\beta=\alpha_1+2\alpha_2+3\alpha_3+4\alpha_4+5\alpha_5+6\alpha_6+7\alpha_7+8\alpha_8
+4\alpha_9+\alpha_{10}+4\alpha_{11}.
\end{align}

As is well-known, this solution is not of finite energy and one has to use the modular $j$-function to `restrict' it to the fundamental domain of the $\tau$ plane such that the energy density becomes proportional to the finite volume of the fundamental domain rather than to the infinite volume of the full upper half plane~\cite{Greene:1989ya,Bergshoeff:2006jj}. One also has to change the conformal factor of the two-dimensional transverse space appropriately. This makes the configuration $SL(2,\ints)$ invariant. If one uses $24$ D$7$ branes the transverse space becomes smooth.

Performing an S-duality on the D$7$ solution produces the S$7$ brane which has the same Einstein frame metric but $\tau$ replaced by $-1/\tau$. On $\beta$ of (\ref{D7root}) it has the effect of increasing the coefficient of $\alpha_{10}$ from $1$ to $3$ and therefore one obtains the root $\beta_1$ of (\ref{d8s3a}).

\end{document}